%% file: main.tex
\documentclass[runningheads]{llncs}
\usepackage[T1]{fontenc}
\usepackage{graphicx}
\usepackage{cite}
\usepackage{amsmath,amssymb,amsfonts}
\usepackage{algorithmic}
\usepackage{graphicx}
\usepackage{textcomp}
\usepackage{xcolor}
\usepackage{array}
\usepackage{xspace}
\usepackage{tabularx}  % Add this in your preamble
\usepackage{booktabs}
\usepackage{xcolor}
\usepackage{multirow}
\usepackage{tcolorbox}
\usepackage{hyperref}
\usepackage[moderate, tracking = normal]{savetrees}
\newcommand{\monitor}{\protect{$\mathcal{M}onitor$}\xspace}
\newcommand{\analyzer}{\protect{$\mathcal{A}nalyzer$}\xspace}
\newcommand{\planner}{\protect{$\mathcal{P}lanner$}\xspace}
\newcommand{\executor}{\protect{$\mathcal{E}xecutor$}\xspace}
\newcommand{\harmone}{\protect{$\mathcal{H}armon{E}$}\xspace}
\def\BibTeX{{\rm B\kern-.05em{\sc i\kern-.025em b}\kern-.08em
    T\kern-.1667em\lower.7ex\hbox{E}\kern-.125emX}}

\makeatletter
\newcommand{\printfnsymbol}[1]{%
  \textsuperscript{\@fnsymbol{#1}}%
}
\makeatother

\begin{document}

\title{HarmonE: A Self-Adaptive Approach to Architecting Sustainable MLOps}

\author{%
  Hiya Bhatt\inst{1}\thanks{Equal contribution.} \and
  Shaunak Biswas\inst{1}\printfnsymbol{1} \and \\
  Srinivasan Rakhunathan\inst{2} \and
  Karthik Vaidhyanathan\inst{1}
}

\institute{%
  Software Engineering Research Centre, IIIT Hyderabad, India\\
  \email{\{hiya.bhatt,shaunak.biswas\}@research.iiit.ac.in}\\
  \email{karthik.vaidhyanathan@iiit.ac.in}
  \and
  Microsoft, India\\
  \email{srrakhun@microsoft.com}
}

\authorrunning{Bhatt et al.}

\maketitle
\begin{abstract}
Machine Learning Enabled Systems (MLS) are becoming integral to real-world applications, but ensuring their sustainable performance over time remains a significant challenge. These systems operate in dynamic environments and face runtime uncertainties like data drift and model degradation, which affect the sustainability of MLS across multiple dimensions: technical, economical, environmental, and social. 
While Machine Learning Operations (MLOps) addresses the technical dimension by streamlining the ML model lifecycle, it overlooks other dimensions. Furthermore, some traditional practices, such as frequent retraining, incur substantial energy and computational overhead, thus amplifying sustainability concerns. To address them, we introduce \harmone, an architectural approach that enables self-adaptive capabilities in MLOps pipelines using the MAPE-K loop. \harmone allows system architects to define explicit sustainability goals and adaptation thresholds at design time, and performs runtime monitoring of key metrics, such as prediction accuracy, energy consumption, and data distribution shifts, to trigger appropriate adaptation strategies. We validate our approach using a Digital Twin (DT) of an Intelligent Transportation System (ITS), focusing on traffic flow prediction as our primary use case. The DT employs time series ML models to simulate real-time traffic and assess various flow scenarios. Our results show that \harmone adapts effectively to evolving conditions while maintaining accuracy and meeting sustainability goals.
\end{abstract}

\input{introduction}

\input{case_study}

\input{approach}

\input{experiment_design}

\input{results}

\input{threats}

\input{related_work}

\input{conclusion}

\input{data_availability}

\input{acknowledgement}
\bibliographystyle{splncs04}
\bibliography{references}





% LNCS Homepage, \url{http://www.springer.com/lncs}, last accessed 2023/10/25

\end{document}

%% file: introduction.tex
\section{Introduction}

Machine Learning Enabled Systems (MLS) are increasingly being deployed in real-world settings such as intelligent transportation~\cite{bhatt_Digit}, healthcare~\cite{healthcare}, and industrial automation~\cite{industrial_automation}. As these systems become integral to everyday applications, ensuring their sustainability has emerged as a critical challenge. Sustainability is recognized as a multi-dimensional quality attribute~\cite{karlskrona}, encompassing \textit{technical} (e.g., model accuracy and maintainability), \textit{environmental} (e.g., energy consumption and emissions), \textit{economic} (e.g., cost-efficiency and resource usage), and \textit{social} (e.g., user trust and long-term impact) dimensions~\cite{lago_sus_dimension}. However, current ML practices often neglect sustainability, given that ML models have evolved to become more computationally intensive over time. In response, the Green AI movement emphasizes resource efficiency and environmental responsibility in AI research and practice~\cite{schwartz2020greenAI}. Despite these efforts, recent surveys indicate that approximately half of ML models fail to transition from prototype to production~\cite{gartner2020}, possibly due to sustainability issues like operational complexity, high resource demands, and long-term maintainability challenges~\cite{maintainability2022slr}. This highlights the importance of explicitly incorporating sustainability into ML deployment practices. Machine Learning Operations (MLOps), which automates the lifecycle of an ML model, has improved technical robustness of MLS but often overlooks environmental and economic impacts. For instance, retraining large models like BERT can consume over 1,500 kWh, emit more than 700 lbs of CO\textsubscript{2}, and cost upwards of \$12,000, highlighting the significant sustainability trade-offs involved \cite{energy_facebook}. Recent literature has considered self-adaptation as a strategy to improve sustainability across all dimensions of MLOps pipelines~\cite{bhatt2024towards}. By enabling systems to autonomously monitor their internal state and external environment, self-adaptive mechanisms can respond to runtime uncertainties such as data drift, performance degradation, or energy overuse. However, existing approaches typically rely on instantaneous reactions, leading to aggressive adaptations that can increase operational overhead \cite{shubham_selfAdap}\cite{tedla2024ecomls}.
\vspace{-4mm}
\begin{figure}[h]
    \centering
    \includegraphics[width=0.7\linewidth]{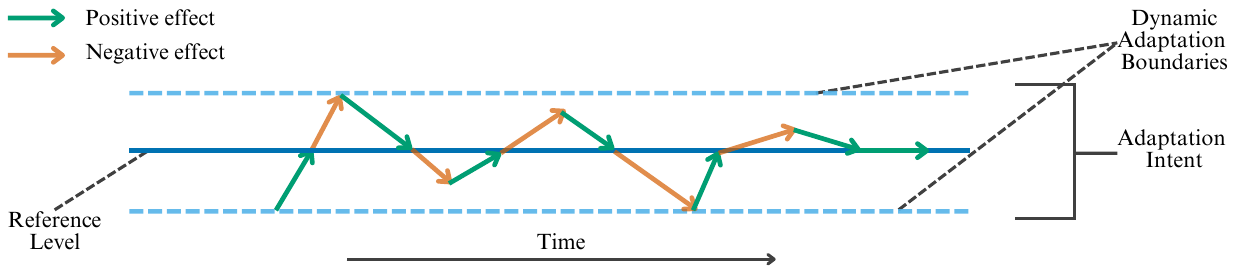}
    \caption{Adaptation explained as a sustainability goal}
    \label{fig:Adaptation_boundaries_dia}
\end{figure}
\vspace{-5mm}

To overcome these limitations, we propose \harmone, an architectural approach that integrates self-adaptive capabilities into MLOps pipelines through the MAPE-K loop \cite{kephart2003vision}, with the goal of supporting long-term sustainability. \harmone continuously monitors metrics such as prediction accuracy, energy consumption, and data distribution shifts, aligning adaptations with sustainability goals established at design time. Prior work has represented adaptation intent as a sustainability goal, defined by maintaining system quality within adaptation boundaries~\cite{adaptation_as_sus_goal_ilias}.
Drawing inspiration from this, we introduce a control-theoretic mechanism centered around a fixed \textit{reference level}, near which the system is expected to operate over time. This enables the system to tolerate temporary deviations, both positive and negative, without triggering immediate adaptation. It allows the \textit{dynamic adaptation boundaries} to evolve gradually based on the cumulative effect of deviations. As illustrated in Figure~\ref{fig:Adaptation_boundaries_dia}, adaptation is only triggered when deviations approach these evolving boundaries, allowing for more deliberate, context-aware responses that ensure the sustainability of the system over time.
We demonstrate \harmone's effectiveness using a traffic flow prediction scenario within a Digital Twin (DT) of an Intelligent Transportation System (ITS). This realistic environment evaluates how \harmone maintains predictive performance and achieves sustainability goals under dynamic real-world conditions. Our results indicate that \harmone effectively balances short-term performance and long-term sustainability objectives across multiple dimensions.

%% file: case_study.tex
\section{Use Case: DT for ITS}
\label{sec:case_study}

As part of an ongoing collaborative effort between \textit{IIIT Hyderabad, India} and \textit{Middlesex University, London} under the \textit{DigIT} project, a DT is being developed for ITS. The DT is intended to support a wide range of urban mobility functionalities, including real-time simulation, monitoring, and control of traffic infrastructure. One of the key system requirements, identified by stakeholders, is the ability to predict the flow of traffic. These predictions will drive downstream tasks such as adaptive traffic re-routing and congestion management within the DT environment. To fulfill this requirement, the system employs an ML-enabled prediction layer integrated into an MLOps pipeline in the DT architecture. The pipeline automates data ingestion from road sensors, preprocessing, model training and validation, deployment, and continuous performance monitoring. Due to the dynamic and non-stationary nature of traffic patterns, which is affected by factors such as time-of-day, accidents, and seasonal variations, the prediction models must be frequently updated to retain accuracy over time. However, this retraining process is computationally intensive, leading to increased energy consumption and associated operational costs. In the early development phase, the Performance Measurement System (PeMS) dataset from California’s highway sensor network was used to benchmark the prediction performance of several ML models under varying conditions.
During this evaluation, the high energy cost and computational overhead of frequent retraining raised concerns about the long-term environmental and economic sustainability of the DT’s predictive capabilities. To address this challenge, we develop \harmone, a self-adaptive framework designed to balance model accuracy with energy efficiency. The PeMS dataset, which includes high-frequency readings from over 40,000 sensors, provides real-world traffic data. For this use case, we use a single sensor node and extract traffic flow aggregated at 5-minute intervals. Predictive models are trained on this data to support traffic forecasting within the DT. This setting enables us to examine how adaptation strategies can contribute to the long-term sustainability of the DT’s predictive layer under real-world variability. We refer to this use case throughout the paper to contextualize \harmone.

%% file: approach.tex
\section{\harmone}
\label{sec:approach}
\harmone is an architectural approach that integrates self-adaptive capabilities into MLOps pipelines to manage runtime uncertainties while aligning with sustainability goals using MAPE-K loop \cite{kephart2003vision}. 
The overall architecture of \harmone is illustrated in Figure~\ref{fig:HarmonE_dia}, which outlines the interaction between its three main components: (1) the \textit{Managed System}, representing the MLS under adaptation; (2) the \textit{Managing System}, which monitors the MLS and its environment, detects uncertainties, and plans and executes adaptations; and (3) the \textit{Decision Map}, which captures sustainability concerns at design time to guide runtime adaptation.
 
\subsection{Managed System}
The \textit{Managed System} encompasses the operational elements of the MLOps pipeline and consists of two key components:

1. \textbf{Training Subsystem}: This subsystem is responsible for training and retraining models when explicitly triggered by the \textit{Managing System} in response to runtime uncertainties such as data/model drift, or sustained performance degradation. After retraining, the new model, along with its corresponding training data, is forwarded to the \textit{Managing System} for versioning.  This ensures that previous models and their corresponding training data are preserved for future reference, reproducibility, and comparison.

% 2. \textbf{Inference Subsystem}: The system relies on a predefined spectrum of models \(\mathcal{M} = \{M_1, M_2, \ldots, M_i\}\), where each model offers a different trade-off between accuracy and resource efficiency. This subsystem is responsible for executing predictions using the currently active model, which is selected and deployed by the Managing System based on runtime conditions and sustainability goals. It does not make selection decisions itself but serves predictions using the model chosen by the Managing System.

2. \textbf{Inference Subsystem}: It performs real-time predictions using a model deployed in the system. The selection of this model is made at runtime by the \textit{Managing System} based on current operating conditions and sustainability goals. 

\begin{figure}[!htbp]
    \centering
    \includegraphics[width=0.9\linewidth]{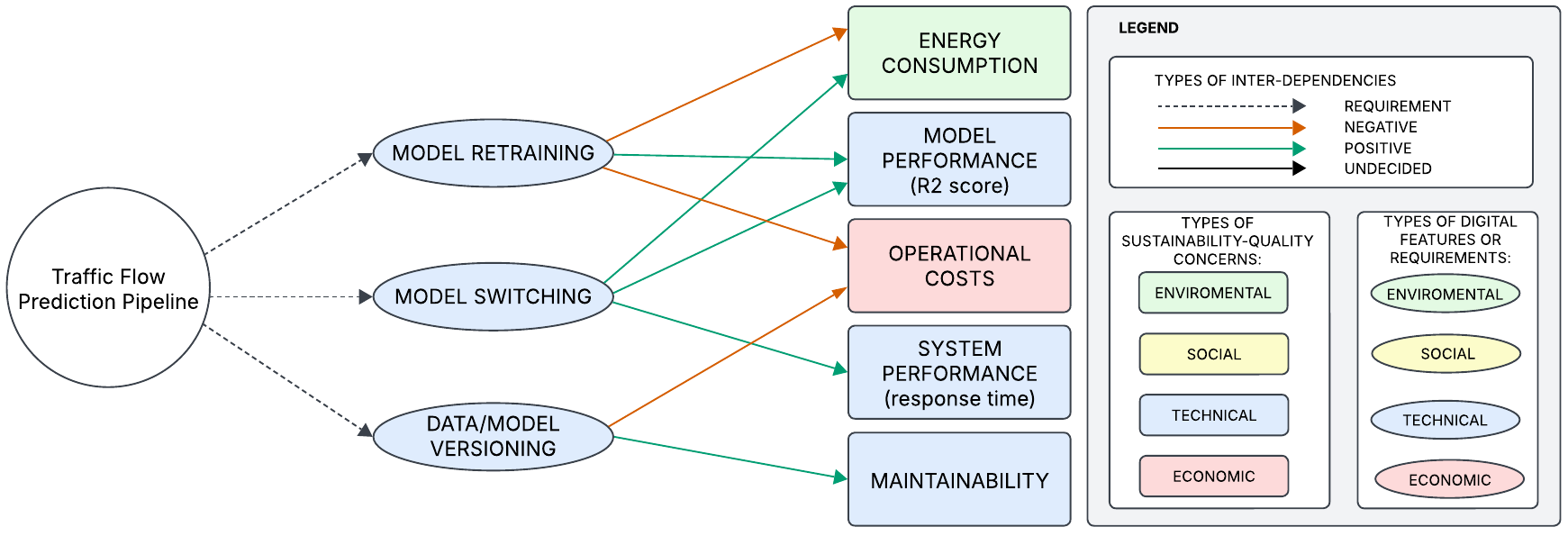}
    \caption{A decision map for a traffic flow prediction pipeline}
    \label{fig:decision-map}
\end{figure}

\subsection{Decision Map}
\label{sec:DM}
A Decision Map (DM), as shown in Figure \ref{fig:decision-map}, is a visual representation of sustainability concerns across the Social, Environmental, Technical and Economic dimensions \cite{lago_sus_dimension}. At design time, the system architect defines key system functionalities, identifies their associated sustainability concerns, and captures their relationships in the DM. Adaptation boundaries are then set based on sustainability goals to guide runtime decisions.

Figure~\ref{fig:decision-map} shows the DM for our DT use case, which is designed for ITS as explained in Section \ref{sec:case_study}. Although the DT encompasses multiple functionalities, such as simulation and monitoring of traffic conditions, this DM specifically addresses sustainability concerns related to the traffic flow prediction component.
For the predictive component of the DT to remain sustainable, it needs to continuously meet its sustainability goals: (a)~Reducing energy consumption; (b)~Improving prediction accuracy and model performance; (c)~Minimizing operational costs; (d)~Enhancing overall system performance; (e)~Ensuring system maintainability. 
To achieve these goals, the DM identifies three functionalities: \textbf{model retraining}, \textbf{data and model versioning}, and \textbf{model switching}. Each functionality affects distinct sustainability dimensions, guiding runtime adaptations through explicit boundaries stored in the \textit{Knowledge Base} of the MAPE-K loop.

\textbf{Model Retraining} happens whenever changes in traffic patterns due to events like road closures or festivals are identified. Retraining models ideally improves predictive accuracy, positively impacting system performance, but also negatively influences environmental and economic sustainability by increasing energy consumption and operational costs (see Figure~\ref{fig:decision-map}). An example of an adaptation boundary here is setting a threshold for allowable data distribution drift (e.g., maximum acceptable KL divergence \cite{kullback1951information}) before retraining is triggered. These boundaries ensure retraining is performed only when necessary, balancing model accuracy against resource utilization.

\textbf{Data and Model Versioning} involves storing trained models along with their corresponding training datasets. This functionality positively affects system maintainability by allowing the reuse of existing models whenever similar traffic conditions reoccur, such as regular rush hours or predictable seasonal variations. However, it negatively impacts operational costs due to additional storage requirements. Hence, an adaptation boundary defined here could be the maximum allowable storage capacity, ensuring the versioning process remains sustainable without excessive overhead.

Lastly, \textbf{Model Switching} allows dynamic selection among pre-trained models based on runtime observations, guided by adaptation boundaries set for metrics like energy consumption and inference accuracy. For instance, in the DT for ITS scenario (refer Section \ref{sec:case_study}), if a highly accurate model surpasses an energy consumption threshold defined in the DM, the system switches to a more energy-efficient model to preserve environmental sustainability. 

% Collectively, these adaptation boundaries, as captured by the DM, enable the system to make informed runtime decisions aligned with predefined sustainability goals, systematically balancing multiple dimensions of sustainability in response to dynamic operational conditions.

\subsection{Managing System}
The \textit{Managing System} is responsible for monitoring the \textit{Managed System} and its environment, detecting uncertainties at runtime, planning, and executing adaptations. It is composed of several components:

\input{knowledge}

\input{monitor}

\input{analyser}

\input{planner}

\input{executor}

%% file: knowledge.tex
\begin{figure}
    \centering
    \includegraphics[width=0.8\linewidth]{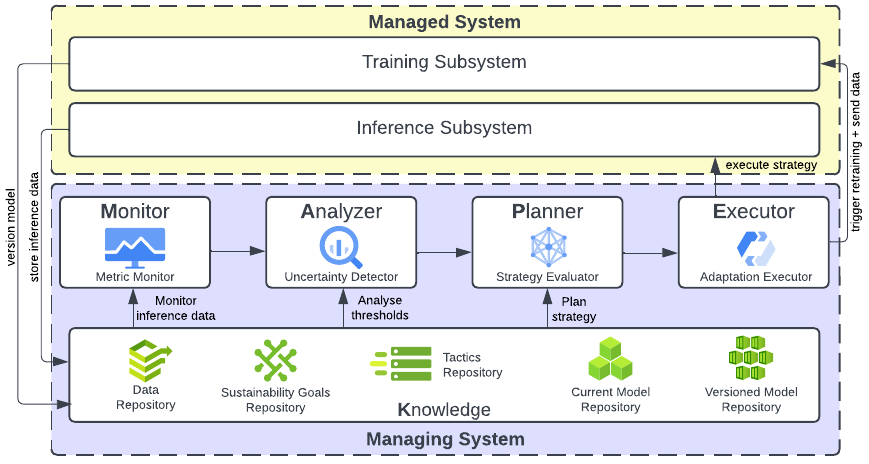}
    \caption{HarmonE Architecture}
    \label{fig:HarmonE_dia}
\end{figure}

1. \textbf{Knowledge Base}:
The \textit{Knowledge Base} serves as the central repository of design-time knowledge, enabling self-adaptive capabilities in the MLS to make informed decisions at runtime. It consists of five key components as shown in Figure~\ref{fig:HarmonE_dia}. First, the \textbf{Data Repository} stores the incoming sensor data along with the corresponding model predictions from the \textit{Inference Subsystem}, enabling drift detection, ongoing performance assessment, and informed model selection during adaptation. Second, the \textbf{Sustainability Goals Repository} stores the sustainability goals defined by the system architect. These goals are translated into measurable thresholds for metrics such as accuracy, energy consumption, and cost. Derived from the system’s DM, these thresholds define the acceptable operational boundaries within which the system is expected to function. Third, the \textbf{Tactics Repository} includes a library of predefined adaptation strategies for managing sustainability trade-offs. During runtime, these strategies are evaluated and selected by the \planner based on the current system state. A detailed list of the tactics used in our implementation is presented in Table~\ref{tab:tactics}. Fourth,
the \textbf{Current Model Repository} maintains a predefined set of \textit{n }models \(\mathcal{M} = \{M_1,~M_2, \ldots, ~M_n\}\) available for deployment during inference. These models are characterized by different trade-offs between accuracy and resource efficiency and are used by the \textit{Managing System} to adaptively select the most suitable model at runtime. Finally, 
the \textbf{Versioned Model Repository}~(VMR) stores retrained versions of models from \(\mathcal{M}\), along with the data distributions on which they were trained. Whenever a model from the current repository is retrained due to drift or performance degradation, the resulting model-data pair is archived here. This enables the system to match incoming data distributions with previously seen ones and reuse past models when appropriate, avoiding redundant retraining and supporting long-term sustainability.

%% file: monitor.tex
2. \textbf{Monitor}:  \monitor consists of a \textit{Metric Monitor} as shown in Figure~\ref{fig:HarmonE_dia}. It observes system behaviour during inference by periodically aggregating runtime metrics at intervals predefined by the system architect. At each interval 
\textit{i}, metrics are aggregated over all new timesteps received since the last monitoring interval. Key metrics include prediction accuracy, energy usage of the system, and data distribution shifts. The specific formulation of prediction accuracy can vary across application domains, for example, $R^2$ score in regression tasks or confidence scores in classification settings. To capture energy efficiency, the system computes a normalized energy value \( \bar{E}_i \) at each interval \( i \), calculated by dividing the current interval's energy consumption by the maximum energy usage observed in the training data. These two metrics are combined into a performance score:
\begin{equation}
S_i = \beta A_i + (1 - \beta)(1 - \bar{E}_i),
\label{eq:tradeoff}
\end{equation}

where \( A_i \) is the accuracy metric (e.g., $R^2$ score), \( \bar{E}_i \) is normalized energy, and \( \beta \in [0,1] \) is a design-time weight balancing accuracy and efficiency. To smooth short-term fluctuations, an Exponential Moving Average (EMA) of \( S_i \) is maintained:   
\begin{equation}
\text{EMA}(S_i) = \gamma S_i + (1 - \gamma) \text{EMA}(S_{i-1})
\label{eq:ema},
\end{equation}

where \( \gamma \) is a smoothing factor that determines the sensitivity to recent performance. To monitor distributional shifts in the data, a suitable divergence measure is used based on the task domain. For instance, in regression settings, KL Divergence \(\mathcal{D}\) is computed between the data distribution of recently observed true values \( P_t \) and a reference distribution \( P_r \), as demonstrated in \cite{augur}. All metrics computed by the \textit{Metric Monitor} are passed to \analyzer as inputs to guide adaptation decisions. For instance, in the DT use case described in Section~\ref{sec:case_study}, these metrics are derived from real-time traffic data captured by a sensor node within California’s highway network.

%% file: analyser.tex
3. \textbf{Analyzer}: \analyzer consists of an \textit{Uncertainty Detector} that evaluates system performance metrics received from \monitor to determine whether runtime adaptation is necessary. Uncertainty is defined as a violation of adaptation boundaries established by the system architect in the DM (Section~\ref{sec:DM}). For performance-based adaptation, \analyzer compares EMA\( (S_i) \), computed by \monitor using Equation~\ref{eq:ema}, to the minimum acceptable threshold \( S_{\min} \) which is stored in the \textit{Knowledge base}. If \( \text{EMA}(S_i) < S_{\min} \), \analyzer identifies a potential performance degradation and triggers \planner to select an appropriate adaptation strategy. For energy efficiency, if the normalized energy consumption \( \bar{E}_i \) at any interval \textit{i} exceeds a dynamic energy threshold $\tau_{E_{i}}$, \analyzer flags this as an uncertainty and calls \planner which triggers adaptation. Inspired by control theory principles \cite{control}, we use the following update rule for $\tau_{E_{i}}$ to adapt to persistent energy deviations:
% \vspace{-0.1mm}
\begin{equation}
    \tau_{E_{i+1}} = \tau_{E_{i}} + \delta \cdot (\bar{E}_{\text{ref}} - \bar{E}_i),
\label{eq:dyn_thresh}
\end{equation}

where \(\bar{E}_{\text{ref}}\) is a design-time reference energy level and \(\delta \in (0,1)\) is a decay factor, both defined in the DM by the system architect. When \(\bar{E}_i > \bar{E}_{\text{ref}}\), the negative deviation \((\bar{E}_{\text{ref}} - \bar{E}_i)\) reduces \(\tau_{E_i}\) to a lower \(\tau_{E_{i+1}}\), tightening the adaptation boundaries to enforce stricter energy constraints. Conversely, if \(\bar{E}_i < \bar{E}_{\text{ref}}\), the positive deviation increases $\tau_{E_{i}}$ to $\tau_{E_{i+1}}$, loosening the adaptation boundaries to accommodate transient energy spikes while maintaining efficiency. For drift-based adaptation, \analyzer evaluates the distribution shift metric \( \mathcal{D} \), computed by \monitor, against a predefined threshold \( \tau_{\text{drift}} \) specified by the system architect. If \( \mathcal{D} > \tau_{\text{drift}} \), an uncertainty is detected, and \planner is invoked to mitigate this drift-related violation. For example, in the DT use case, a sudden change in traffic patterns may increase \( \mathcal{D} \) beyond the acceptable limit, prompting \analyzer to initiate appropriate drift-handling adaptations.

%% file: planner.tex
\input{uncertainties_table}
4. \textbf{Planner}: \planner consists of a \textit{Strategy Evaluator}, as shown in Figure \ref{fig:HarmonE_dia}, which is responsible for selecting and initiating the appropriate adaptation strategy whenever \analyzer detects a runtime uncertainty. These strategies are derived from the set of predefined tactics \cite{leest2023evolvability},\cite{green_tactics_lago} associated with each type of uncertainty, as outlined in Table~\ref{tab:tactics}, and are stored in the Tactics Repository within the \textit{Knowledge Base} (see Figure~\ref{fig:HarmonE_dia}). The table also categorizes the impact of each uncertainty as \textit{immediate} (direct operational effects), \textit{enabling} (mid-term consequences on resource use or retraining), and \textit{systemic} (long-term or structural implications) \cite{lago_sus_dimension}. Thus, the \textit{Strategy Evaluator} makes decisions based on the current runtime context and accumulated historical performance metrics. For performance-based adaptation, the \textit{Strategy Evaluator} retrieves the model-specific EMA scores from the \textit{Knowledge Base} and selects from the spectrum of models, excluding the currently active one, the model that yields the highest EMA\( (S_i) \) (refer to Equation~\ref{eq:ema}). An \(\epsilon\)-greedy exploration strategy is used, wherein with probability \(\epsilon\), the \textit{Strategy Evaluator} randomly selects a model from the \textit{current model repository} \(\mathcal{M}\) in the Knowledge Base. For energy-based adaptation, if the normalized energy consumption \(\bar{E}_i\) for interval \(i\) exceeds the dynamic threshold \(\tau_{E_i}\), \planner is triggered to reassess the deployed model. When an energy violation occurs, the \textit{Strategy Evaluator}  selects the model with the best EMA\( (S_i) \) among those excluding the currently active model, thereby addressing environmental sustainability concerns without compromising predictive performance.  In case of drift-based adaptation, if the distribution shift metric \(\mathcal{D}\) exceeds the drift threshold \(\tau_{\text{drift}}\), the planner compares the current data distribution with the stored training distributions of versioned models stored in the VMR in the \textit{Knowledge Base} (see Figure \ref{fig:HarmonE_dia}). If a suitable match is found, the system switches to that versioned model; otherwise, \executor is triggered for full retraining. This planning mechanism allows the system to adapt dynamically to runtime uncertainties while preserving long-term sustainability goals. For instance, in the DT use case discussed in Section~\ref{sec:case_study}, if a sudden traffic disruption leads to reduced accuracy or increased energy usage, the planner may decide to switch from an LSTM model to a lighter linear model until stability is restored.

%% file: uncertainties_table.tex
\begin{table*}[h]
\centering
\scriptsize % Reduce font size
% \fontsize\{1pt\}{1pt\selectfont}
\caption{Self-Adaptation Uncertainties, Tactics, and Strategies for Sustainable MLOps}

\begin{tabular}{|>{\raggedright\arraybackslash}p{1.6cm}|>{\raggedright\arraybackslash}p{1.7cm}|>{\raggedright\arraybackslash}p{1.4cm}|>{\raggedright\arraybackslash}p{1.7cm}|>{\raggedright\arraybackslash}p{1.2cm}|>{\raggedright\arraybackslash}p{1.8cm}|>{\raggedright\arraybackslash}p{2.0cm}|}
\hline
\textbf{Uncertainty} & \textbf{Concern} & \multicolumn{3}{c|}{\textbf{Impact}} & \textbf{Tactics} & \textbf{Strategy} \\ \cline{3-5}
& & \textbf{Immediate} & \textbf{Enabling} & \textbf{Systemic} & & \\ \hline

\textbf{1. Model Drift} & Technical & Reduced prediction quality & Reduced user trust & Reduced social benefits & \textbf{1.~Model switch} \newline \newline \textbf{2. Retrain}& a. Switch to better-fit model from VMR \newline b. Retrain via Incremental/Transfer Learning/Full retraining \\ \hline

\textbf{2. Data Drift} & Technical & Degraded due to data shifts & Inaccurate predictions & Trust loss, low relevance & \textbf{1.~Model switch} \newline \newline \textbf{2. Retrain} & a. Switch to better-fit model from VMR \newline b. Retrain with resampled/augmented data \\ \hline

\textbf{3. High Energy Consumption} & Environmental & Increased costs & Environmental impact & Lower competitiveness & \textbf{1. Model switch \newline 2. Model Compression} & a. Switch to lighter model \newline b. Quantize current model \\ \hline

\textbf{4. Model performance degradation} & Technical & Decline in predictive accuracy & Increased need for \textit{frequent} retraining & Reduced maintainability & \textbf{1.~Model switch} \newline \newline \textbf{2. Retrain} & a. Switch to a model with higher accuracy \newline b. Retrain the model  \\ \hline

\end{tabular}

\label{tab:tactics}
\end{table*}

%% file: executor.tex
5. \textbf{Executor}:  
\executor is responsible for enacting the adaptation strategies determined by the \planner. When the selected strategy involves retraining, \executor calls the \textit{Training Subsystem} to initiate the process. Once training is complete, \executor versions the trained model and its associated data in the VMR in the \textit{Knowledge Base}. If the strategy involves switching to a different model, either from the VMR or from the predefined spectrum of models \(\mathcal{M}\), \executor updates the model used by the \textit{Inference Subsystem} accordingly. In doing so, it ensures that subsequent predictions are made using the model selected by the \planner.

%% file: experiment_design.tex
\section{Experiment Design}

% We assess the efficiency of our approach by answering the following Research Questions~(RQ):
We assess the efficiency of \harmone by answering these Research Questions~(RQ):

\noindent\textbf{RQ1}: How does \harmone ensure the long-term environmental sustainability of MLS?

\noindent\textbf{RQ2}: How does \harmone manage the trade-off between energy consumption and predictive accuracy compared to baseline approaches?

\noindent\textbf{RQ3}: How efficient are the adaptation decisions made by \harmone in terms of their frequency and resource usage?

The remainder of this section describes the setup and design of the experiments to evaluate our system. 
\vspace{-4mm}
\subsection{Experimental Setup}
\label{sec:exp_set}

To evaluate \textit{\harmone}, we design an experimental setup that reflects the challenges and operational dynamics of sustainability-aware inference in real-world systems. As described in the use case (Section~\ref{sec:case_study}), we utilize traffic flow (\texttt{Vehicles/5 Minutes}) data collected from sensor nodes deployed as part of the California PeMS platform. 

%These sensors capture information such as vehicle speed and traffic flow at 5-minute intervals.

For forecasting, we structure the data in a supervised learning format where each model receives the previous 5 timesteps (i.e., 25 minutes of sensor readings) as input to predict the traffic flow for the 6th timestep (i.e., the 30th minute). This setup allows us to evaluate the models' ability to learn short-term temporal dependencies while operating under energy and latency constraints. Further, we construct a spectrum of models that vary in both predictive accuracy and computational cost:

\begin{itemize}
    \item \textbf{Linear Regression (LR)}: Lightweight and fast but less expressive. It serves as the most energy-efficient model \cite{data_green_roberto}.
    \item \textbf{Support Vector Machine (SVM)}: Balances moderate accuracy with intermediate inference time and energy consumption.
    \item \textbf{Long Short-Term Memory (LSTM)}: Offers highest accuracy but comes with the highest energy and latency overhead.
\end{itemize}

% While LSTM models often outperform others in terms of accuracy, we observe that in several intervals, LR and SVM achieve satisfactory performance at a fraction of the energy cost. This justifies their presence in the inference subsystem despite their lower average accuracy.
% The training phase uses a small chunk of historical data from the selected sensor node. Let \( r \) represent the number of rows in the training window, which is fixed across all models to ensure uniform comparison. After training, the models are deployed in the inference subsystem, as shown in the system architecture (Figure~\ref{fig:HarmonE_dia}). Inference is then carried out in real-time on incoming data from the sensor. During inference, the system logs key metrics, including the predicted value, the ground-truth value, the time taken for prediction, and the energy consumed.
We conducted a pilot experiment over a short interval and observed that while LSTM generally provides higher accuracy, LR and SVM models often yield satisfactory performance at significantly lower energy cost. This supports their inclusion in the inference setup to enable more energy-efficient predictions when appropriate. All models are trained on a fixed-size historical dataset to ensure uniform comparison. After training, they are deployed in the \textit{Inference Subsystem} for real-time inference on incoming sensor data. During inference, the system logs key metrics such as prediction output, ground truth, inference time, and energy consumption.
The coefficient of determination \(R^2\) is used to evaluate predictive accuracy. Energy consumption is measured using the \texttt{pyRAPL}\footnote{https://pypi.org/project/pyRAPL/ (last accessed May 20, 2025)} library, which provides high-resolution energy profiling at the system level. The models were implemented using the Python libraries \texttt{scikit-learn} for Linear Regression and SVM, and \texttt{PyTorch} for LSTM. For details of the implementation, please refer to our Github repository \footnote{https://github.com/sa4s-serc/HarmonE (last accessed May 20, 2025)}.

\subsection{Baselines Setup}

To systematically evaluate the sustainability and effectiveness of \harmone, we compare it against the following baselines:

\begin{itemize}
    \item \textbf{LR}: Inference using Linear Regression model
    \item \textbf{LR+PRT}: Linear Regression with periodic retraining.
    \item \textbf{SVM}: Inference using Support Vector Machine.
    \item \textbf{SVM+PRT}: Support Vector Machine with periodic retraining.
    
    \item \textbf{LSTM}: Inference using Long Short-Term Memory.
    \item \textbf{LSTM+PRT}: Long Short-Term Memory with periodic retraining.
    
    \item \textbf{Switch}: Switching among LR, SVM, and LSTM based on uncertainties detected by the Managing System (see Section~\ref{sec:approach}); however, no retraining is performed.
    \item \textbf{Switch+PRT}: \textbf{Switch} with periodic retraining. 
    
    \item \textbf{\harmone}: Our approach, as explained in Section \ref{sec:approach}
\end{itemize}

We divide the dataset into three parts: 80\% (1200 samples) is used for training, 20\% (240 samples) for validation, and a separate test set of 14,500 samples is used for inference and evaluation. All models are trained using the same training-validation split to ensure consistency across approaches. Retraining, whether periodic or triggered (\harmone), is performed using 1200 samples, consistent with the original training window size, with periodic retraining triggered uniformly every 3200 timesteps. To ensure consistent and reliable results, each baseline experiment is repeated independently five times, following established guidelines for empirical software engineering \cite{wohlin2012experimentation}. Additionally, a cooldown period of 20 minutes is introduced between consecutive runs to stabilize hardware conditions and ensure accurate energy profiling. All experiments are conducted in a controlled simulation environment, enabling reproducibility and fair comparisons. The performance of these approaches is summarized in Table~\ref{tab:results}.

\subsection{Drift Induction}

To evaluate \harmone's performance in the presence of real-world uncertainties, we simulate distribution drift in the test dataset. This is done to mimic scenarios where external factors, such as traffic rerouting due to roadblocks or unexpected events like accidents, cause abrupt changes in data patterns. Specifically, we induce controlled data drift by applying a consistent scale-and-shift transformation twice to designated segments of the test data. Our drift induction method, inspired by \cite{augur}, ensures the changes are measurable and representative of natural concept drift, thereby enabling meaningful evaluation of \harmone’s adaptation capabilities. By repeating the same transformation, we simulate a scenario where the data distribution first diverges and later realigns with a previously seen pattern. This enables \harmone to demonstrate both drift detection and effective model reuse by identifying similarity with historical distributions stored in the VMR.

\subsection{System Specifications}
All experiments were conducted on a system with the following specifications: Debian GNU/Linux 12 (Bookworm) operating system with kernel version 6.1.0-32-amd64, running on a 12th Gen Intel i5-1240P processor at 4.40 GHz. The system had 8 GB RAM, an integrated Intel Alder Lake-P GPU, and utilized the GNOME 43.9 desktop environment. The consistent hardware configuration ensured fair comparisons and reproducible energy profiling across all experiments.

%% file: results.tex
\section{Results}

\vspace{-7mm}
\begin{figure}[h]
\centering
\includegraphics[width=\textwidth]{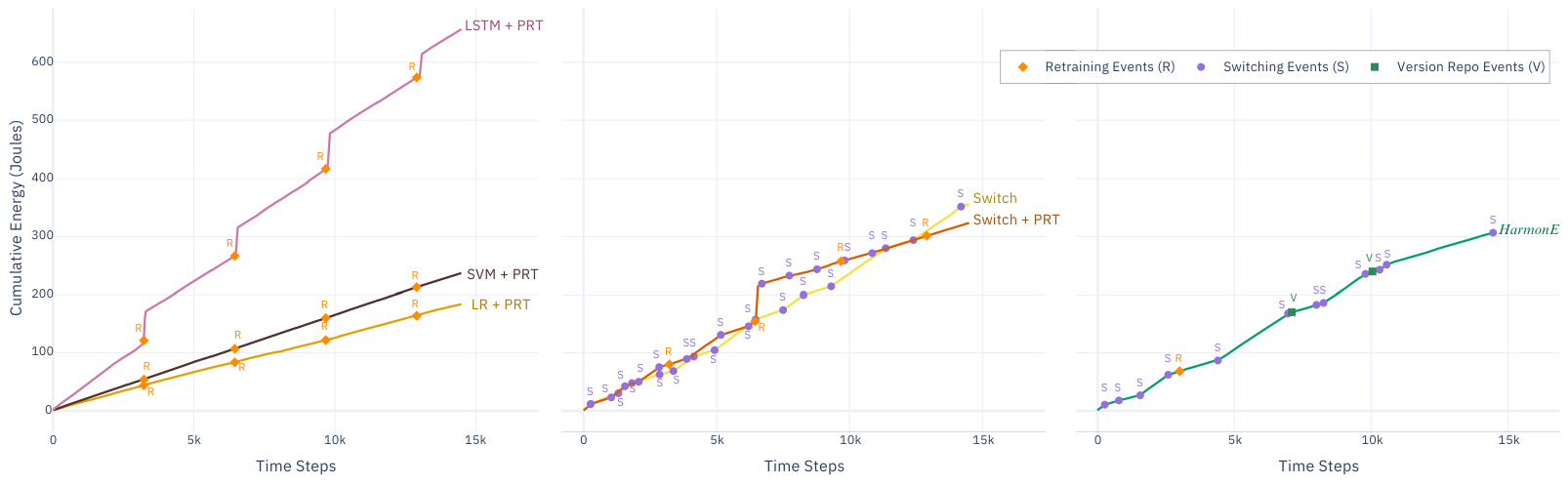}
\caption{Adaptation strategies activated by \harmone during execution.}
\label{fig:adaptation}
\end{figure}

\begin{figure}[h]
\centering
\includegraphics[width=0.85\textwidth]{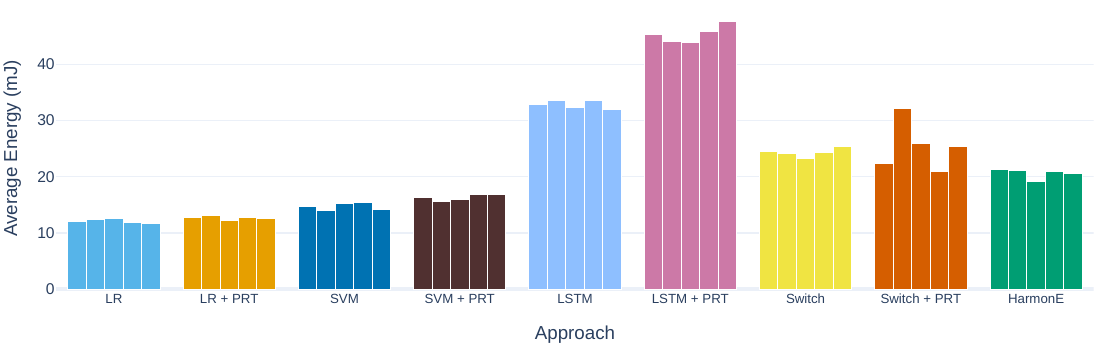}
\caption{Average energy consumption (mJ) across five runs for all nine approaches.}
\label{fig:bar_chart}
\end{figure}

\vspace{-5mm}
This section presents the experimental results of our proposed approach, \harmone, in comparison to eight baseline approaches. Figure~\ref{fig:adaptation} illustrates the adaptation strategies triggered by \harmone throughout execution. The temporal distribution of switching (S), retraining (R), and version repository access (V) events reflects how \harmone responds to evolving conditions while balancing adaptation costs. To ensure reliability, we conducted five independent runs for each approach and averaged the energy consumption. As shown in Figure~\ref{fig:bar_chart}, results are consistent across runs, with minor variation in Switch+PRT due to stochastic model selection during retraining. Despite this, the relative performance trends remain stable, supporting the reproducibility of our findings.
% \vspace{-5mm}

% \vspace{-5mm}
\vspace{-7mm}
\textbf{RQ1: How does \harmone ensure the long-term environmental sustainability of MLS?}

\begin{tcolorbox}[
    colback=green!5!white, 
    colframe=green!40!black,
    boxrule=0.5pt,
    arc=1pt,
    boxsep=2pt,
    left=4pt,
    right=4pt,
    top=2pt,
    bottom=2pt
]

\harmone ensures long-term sustainability by maintaining energy consumption below threshold constraints. It consumes \textbf{54.5\%} lesser energy as compared to high-performing LSTM+PRT.

\end{tcolorbox}

\begin{figure}[h]
\centering
\includegraphics[width=0.86\textwidth]{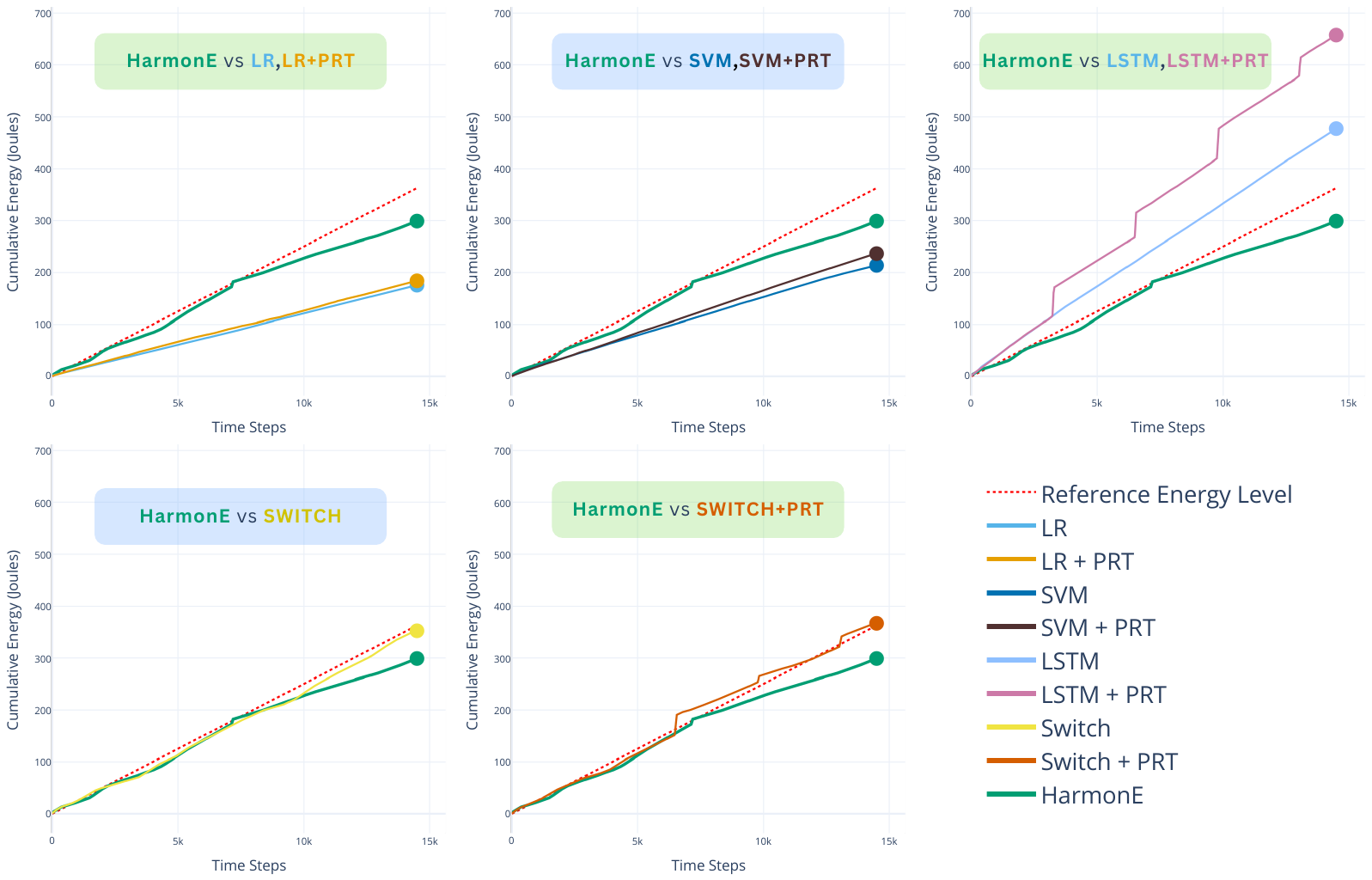}
\caption{Cumulative energy consumption over time for different approaches.}
\label{fig:cumulative_energy}
\end{figure}
\vspace{-5mm}
To analyze environmental sustainability, we examine cumulative energy consumption over time steps (Figure~\ref{fig:cumulative_energy}) for all approaches. \harmone consistently maintains its energy consumption below the reference energy level (red dotted line) defined by the system architect at design time using the DM (refer \ref{sec:DM}), by dynamically adapting whenever this boundary is violated. Comparatively, approaches like LR, LR+PRT, SVM, and SVM+PRT show consistently low energy use but significantly underperform in predictive accuracy (discussed further in the section). In contrast, approaches like LSTM, LSTM+PRT significantly surpass \harmone~’s energy consumption. Approaches like Switch, Switch+PRT perform better than single models but still exhibit higher energy consumption than \harmone. It is observed that all the approaches with PRT, retrain models after fixed intervals regardless of necessity thus consuming more energy, whereas \harmone only retrains the models whenever drift is detected (refer Section \ref{sec:approach}) thus significantly reducing the energy consumption. Quantitatively, \harmone consumes 20.62 mJ, representing a 54.5\% reduction compared to LSTM+PRT (45.35 mJ) and an 18.6\% reduction compared to Switching+PRT (25.32 mJ).

\textbf{RQ2: How does \textbf{\harmone} manage the trade-off between energy consumption and predictive accuracy compared to baseline approaches?}

\begin{tcolorbox}[
    colback=green!5!white, 
    colframe=green!40!black,
    boxrule=0.5pt,
    arc=1pt,
    boxsep=2pt,
    left=4pt,
    right=4pt,
    top=2pt,
    bottom=2pt
]
\harmone effectively manages the trade-off between energy consumption and predictive accuracy. It achieves \textbf{95.0\%} of LSTM+PRT's accuracy while consuming only \textbf{45.5\%} of its energy and improves inference time by \textbf{53.4\%}.
\end{tcolorbox}

\begin{figure}[h]
\centering
\includegraphics[width=\textwidth]{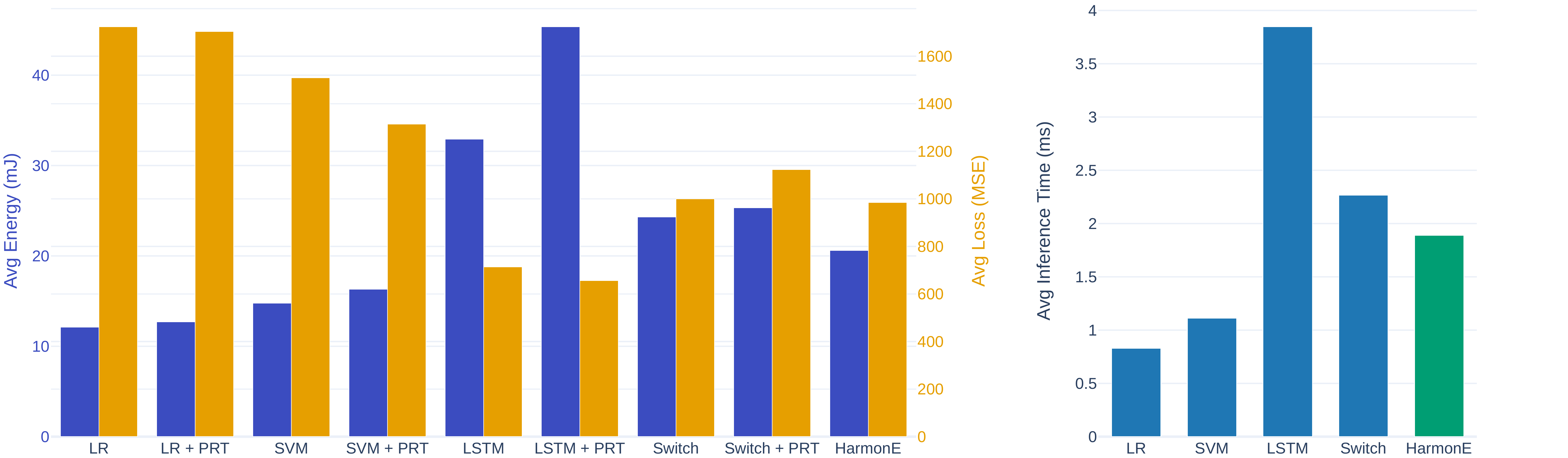}
\caption{Energy vs. prediction error (MSE) across approaches—lower is better on both axes~(on left). Inference time comparison excluding retraining—lower is better~ (on right).}
\label{fig:energy_loss}
\end{figure}
To address RQ2, we analyze how different approaches balance predictive accuracy, energy consumption, and inference speed using the results summarized in Figure~\ref{fig:energy_loss} and Table~\ref{tab:results}. LR and SVM-based approaches achieve minimal energy consumption (12.1157–16.3304 mJ, Table~\ref{tab:results}), but exhibit significantly higher prediction errors (1315-1723 MSE), Figure~\ref{fig:energy_loss}) and lower predictive accuracy ($R^2$ scores ranging from 0.7597 to 0.8167). Conversely, LSTM-based approaches deliver lower prediction errors (657–713 MSE) and higher predictive accuracy ($R^2$ = 0.9005 and 0.9085) at the expense of substantial energy consumption (32.9147–45.3517 mJ) and inference latency (3.8475 ms). Periodic retraining, irrespective of necessity, further escalates these energy and latency costs. The Switch approach, despite aiming for balance, defaults to the heavier LSTM model without retraining smaller models when drift occurs, resulting in higher average energy usage (24.3218 mJ) without achieving LSTM-level accuracy ($R^2$ = 0.8445). In contrast, \harmone dynamically manages model selection, selective retraining, and versioned model reuse, achieving an $R^2$ score of 0.8627—95.0\% of LSTM+PRT accuracy—at only 45.5\% of its energy cost. Additionally, \harmone attains faster inference times (1.89 ms), reducing latency by 53.4\% compared to LSTM+PRT (4.05 ms). These results highlight \harmone's ability to efficiently balance energy and accuracy, significantly outperforming baseline approaches.

\vspace{-5mm}
\input{results_table}

\vspace{-5mm}
\textbf{RQ3: How efficient are the adaptation decisions made by \harmone in terms of their frequency and resource usage?}

\begin{tcolorbox}[
    colback=green!5!white, 
    colframe=green!40!black,
    boxrule=0.5pt,
    arc=1pt,
    boxsep=2pt,
    left=4pt,
    right=4pt,
    top=2pt,
    bottom=2pt
]
\harmone makes highly efficient adaptation decisions, with the decision-making process itself consuming only 1.3660\% of total system energy while enabling strategic resource optimization through context-aware adaptation mechanisms.
\end{tcolorbox}

To address RQ3, we analyze the frequency of adaptation decisions and their resource consumption. Our experiments reveal that \harmone's adaptation decisions are highly efficient in terms of both frequency and resource usage.  The adaptation logic (MAPE-K loop) executes swiftly, averaging 17.899 ms per invocation, thereby imposing negligible runtime overhead. The energy consumed by the MAPE-K loop accounts for only 1.3660\% of the total energy usage of the system. This extremely low overhead demonstrates that \harmone's decision-making mechanism itself is highly energy-efficient. Periodic retraining approaches (LR+PRT, SVM+PRT, LSTM+PRT, Switch+PRT) conduct retraining at fixed intervals (4 per run), regardless of necessity, leading to redundant resource usage. In contrast, purely switching-based adaptation (Switch) triggers frequent model switches (average of 13 per run) without retraining, often defaulting to heavier models due to degraded performance in lighter models. \harmone addresses these inefficiencies by selectively performing adaptations, either switching, retraining, or reusing versioned models, only when required by system conditions (as shown in Figure~\ref{fig:adaptation}). This significantly reduces unnecessary resource consumption while maintaining predictive performance.

%% file: results_table.tex
\begin{table}[hbtp]
\scriptsize
\centering
\caption{Performance comparison of different approaches.}
\label{tab:results}
\begin{tabular}{l c c c c c}
\toprule
\textbf{Approach} & \textbf{Energy(mJ)} & \textbf{R\textsuperscript{2}} & \textbf{Inference Time(ms)} & \textbf{\# Adaptations} \\
\midrule
LR                    & 12.1157 & 0.7597 & \multirow{2}{*}{0.8306} & -  \\
LR + PRT          & 12.6987 & 0.7624 &                        & 4 \\
\midrule
SVM                   & 14.7649 & 0.7897 & \multirow{2}{*}{1.1134} & - \\
SVM + PRT         & 16.3304 & 0.8167 &                        & 4  \\
\midrule
LSTM                  & 32.9147 & 0.9005 & \multirow{2}{*}{3.8475} & -  \\
LSTM + PRT        & 45.3517 & 0.9085 &                       & 4  \\
\midrule
Switch             & 24.3218 &  0.8445 & \multirow{2}{*}{2.2656} & 13 \\
Switch + PRT   & 25.3176 &  0.8435 &                       & 17  \\
\midrule
\textbf{\harmone} & \textbf{20.6203} & \textbf{0.8628} & \textbf{1.8887} & \textbf{12} \\
\bottomrule
\end{tabular}
\end{table}

%% file: threats.tex
\section{Threats to Validity}
A threat to \textit{internal validity} arises from the consistency of results across repeated experimental runs, as energy consumption and performance metrics can be affected by background processes and runtime fluctuations. To mitigate this, all experiments were repeated independently five times for each approach, and average results were reported to ensure consistency. Additionally, a cooldown period of 20 minutes was enforced between consecutive runs to allow system conditions, particularly energy usage and temperature, to stabilize before the next experiment. A threat to \textit{Construct Validity} could be that the energy measurements in our experiments were collected using the \texttt{pyRAPL} library, Python package specifically designed for assessing the energy consumption and power usage of software applications running on Intel processors. While the absolute energy values may differ when using alternative measurement tools or hardware platforms, we believe that the relative patterns and trends in adaptation behaviour would remain consistent. 
A threat to \textit{external validity} concerns the generalizability of our findings beyond the specific use case of traffic flow prediction. Although this represents a realistic and dynamic application, it may not fully reflect the characteristics of other domains such as computer vision or natural language processing. Additionally, our evaluation focuses on a single task type and a limited set of models. To mitigate this, we employed a spectrum of models with varying complexity and resource requirements to simulate diverse operational conditions. While further validation is needed in other domains, \harmone is designed to be domain-agnostic and can be extended to different MLS through appropriate domain knowledge.
% \textbf{External Validity:} The experimental setup is based on traffic flow prediction from a single sensor node in the PeMS dataset. While this use case reflects a realistic and dynamic environment with clear sustainability challenges, the evaluation may not generalize directly to domains such as computer vision or NLP, where model characteristics and resource profiles differ significantly. Moreover, our current scope is limited to CPU-bound execution; evaluating \harmone in distributed or GPU-accelerated MLOps pipelines remains future work. Nonetheless, the architecture of \harmone—goal-driven adaptation, reference thresholding, and sustainability-aware planning—are designed to be domain-agnostic and can be applied to other settings with appropriate metric definitions and tactics.

%% file: related_work.tex
\section{Related Work}

There has been substantial progress in developing MLOps frameworks to automate the ML model lifecycle, as seen in works such as \cite{symeonidis2022mlops,shankar2020operationalizing,amershi2019software}, but these primarily emphasize maintainability, scalability, and reliability, without addressing sustainability as an architectural concern. Notable efforts have been made to employ self-adaptation in MLS, such as \cite{kulkarni2023towards}, which proposes dynamic model switching strategies using MAPE-K loops to balance Quality of Service~(QoS) metrics at runtime, yet remain limited to inference and do not incorporate energy, retraining cost, or long-term sustainability trade-offs. EcoMLS \cite{tedla2024ecomls} introduces energy-aware switching during inference, marking a shift toward integrating sustainability at runtime, but it is too restricted in scope and does not address sustainability across the full ML lifecycle, including training and continuous evolution. Work in software sustainability, including \cite{lago_sus_dimension,adaptation_as_sus_goal_ilias}, advocates architectural DM and the expression of adaptation intent as sustainability goals, but lacks integration with runtime MLOps adaptation to handle uncertainties typical in MLS. In parallel, the ``Green AI'' community focuses on efficient model architectures \cite{han2016deepcompression,cai2019onceforall}, offering energy savings at the algorithmic level, but these methods are typically static and disconnected from lifecycle or runtime MLOps concerns. Work such as \cite{tamburri2020sustainablemlops} articulates the organizational and architectural challenges in achieving sustainable MLOps, while others like \cite{leest2023evolvability} explore architectural tactics to support evolvability in ML pipelines, though without mechanisms to operationalize these decisions adaptively. \cite{mlsenvironment} highlights the absence of integrated, end-to-end architectural support for environmental sustainability in ML system engineering through a systematic mapping study. While these efforts lay essential groundwork, a lack of approaches that operationalize sustainability across the ML lifecycle by adaptively managing runtime uncertainties and evolving trade-offs remains. To address this, we present \harmone, an architectural approach that enables self-adaptive capabilities in MLOps pipelines via the MAPE-K loop. \harmone dynamically balances predictive performance and energy consumption throughout the ML lifecycle.

%% file: conclusion.tex
\section{Conclusion}
This paper presented \harmone, a self-adaptive architectural approach that integrates the MAPE-K loop within MLOps pipelines to address long-term sustainability challenges in MLS deployments. \harmone dynamically adjusts system configurations through strategies such as model switching, selective retraining, and versioned model reuse, rather than relying on fixed-interval retraining. These design choices allow the system to continuously balance energy consumption and predictive accuracy. Experimental results obtained using a DT of an ITS demonstrate that \harmone maintains environmental sustainability over extended periods. Specifically, \harmone reduces energy consumption by 54.5\% compared to LSTM+PRT while preserving 95\% of its predictive accuracy. The approach outperforms periodic retraining methods by triggering adaptations only when sustainability boundaries are reached, thereby reducing unnecessary resource expenditure and mitigating long-term operational overhead. Future work will focus on validating \harmone in more diverse and complex real-world environments, including domains such as computer vision and natural language processing. In particular, we aim to evaluate its applicability in LLM/SLM-enabled systems where the sustainability trade-offs of model selection, fine-tuning, and inference are more pronounced, further extending its relevance to a broader class of MLS deployments.

%% file: data_availability.tex
\section{Data Availability}
The traffic flow data used in this study is publicly accessible via the California Performance Measurement System (PeMS) website at \href{https://pems.dot.ca.gov/}{https://pems.dot.ca.gov}. For reproducibility and verifiability, the source code and configurations for implementing \harmone are available at \href{https://github.com/sa4s-serc/HarmonE}{https://github.com/sa4s-serc/HarmonE}. 
% For further assistance, contact the corresponding authors.

%% file: acknowledgement.tex
\section{Acknowledgements}

The authors would like to thank the ANRF Prime Minister Early Career Research Grant (ANRF/ECRG/2024/003379/ENS) and the UKIERI-SPARC project titled “DigIT—Digital Twins for Integrated Transportation Platform” (UKIERI-SPARC/01/23) for their support.

%% file: main.bbl
\begin{thebibliography}{10}
\providecommand{\url}[1]{\texttt{#1}}
\providecommand{\urlprefix}{URL }
\providecommand{\doi}[1]{https://doi.org/#1}

\bibitem{amershi2019software}
Amershi, S., Begel, A., Bird, C., DeLine, R., Gall, H., Kamar, E., Nagappan, N., Nushi, B., Zimmermann, T.: Software engineering for machine learning: A case study. In: 2019 IEEE/ACM 41st International Conference on Software Engineering: Software Engineering in Practice (ICSE-SEIP) (2019)

\bibitem{karlskrona}
Becker, C., Chitchyan, R., Duboc, L., Easterbrook, S., Penzenstadler, B., Seyff, N., Venters, C.C.: Sustainability design and software: The karlskrona manifesto. In: 2015 IEEE/ACM 37th IEEE International Conference on Software Engineering. vol.~2 (2015)

\bibitem{bhatt2024towards}
Bhatt, H., Arun, S., Kakran, A., Vaidhyanathan, K.: Towards architecting sustainable mlops: A self-adaptation approach. In: ICSA-C (2024)

\bibitem{bhatt_Digit}
Bhatt, H., Vaidhyanathan, K., Biju, R., Gangadharan, D., Trestian, R., Shah, P., et~al.: Architecting digital twins for intelligent transportation systems. arXiv preprint arXiv:2502.17646  (2025)

\bibitem{cai2019onceforall}
Cai, H., Gan, C., Wang, T., Zhang, Z., Han, S.: Once-for-all: Train one network and specialize it for efficient deployment. arXiv preprint arXiv:1908.09791  (2019)

\bibitem{mlsenvironment}
Chadli, K., Botterweck, G., Saber, T.: The environmental cost of engineering machine learning-enabled systems: A mapping study. In: Proceedings of the 4th Workshop on Machine Learning and Systems. EuroMLSys '24, Association for Computing Machinery, New York, NY, USA (2024)

\bibitem{gartner2020}
Costello, K., Rimol, M.: Gartner identifies the top strategic technology trends for 2021

\bibitem{control}
Franklin, G., Powell, J., Workman, M.: Digital Control of Dynamic Systems-Third Edition (2022)

\bibitem{adaptation_as_sus_goal_ilias}
Gerostathopoulos, I., Raibulet, C., Lago, P.: Expressing the adaptation intent as a sustainability goal. In: Proceedings of the ACM/IEEE 44th International Conference on Software Engineering: New Ideas and Emerging Results (2022)

\bibitem{han2016deepcompression}
Han, S., Mao, H., Dally, W.J.: Deep compression: Compressing deep neural networks with pruning, trained quantization and huffman coding. arXiv preprint arXiv:1510.00149  (2015)

\bibitem{green_tactics_lago}
J\"{a}rvenp\"{a}\"{a}, H., Lago, P., Bogner, J., Lewis, G., Muccini, H., Ozkaya, I.: A synthesis of green architectural tactics for ml-enabled systems. In: Proceedings of the 46th International Conference on Software Engineering: Software Engineering in Society. ICSE-SEIS'24, Association for Computing Machinery, New York, NY, USA (2024)

\bibitem{kephart2003vision}
Kephart, J.O., Chess, D.M.: The vision of autonomic computing. Computer  \textbf{36}(1) (2003)

\bibitem{shubham_selfAdap}
Kulkarni, S., Marda, A., Vaidhyanathan, K.: Towards self-adaptive machine learning-enabled systems through qos-aware model switching. In: 2023 38th IEEE/ACM International Conference on Automated Software Engineering (ASE). IEEE (2023)

\bibitem{kulkarni2023towards}
Kulkarni, S., Marda, A., Vaidhyanathan, K.: Towards self-adaptive machine learning-enabled systems through qos-aware model switching. In: 2023 38th IEEE/ACM International Conference on Automated Software Engineering (ASE) (2023)

\bibitem{kullback1951information}
Kullback, S., Leibler, R.A.: On information and sufficiency. The annals of mathematical statistics  \textbf{22}(1) (1951)

\bibitem{lago_sus_dimension}
Lago, P.: Architecture design decision maps for software sustainability. In: 2019 IEEE/ACM 41st International Conference on Software Engineering: Software Engineering in Society (ICSE-SEIS). IEEE (2019)

\bibitem{leest2023evolvability}
Leest, J., Gerostathopoulos, I., Raibulet, C.: Evolvability of machine learning-based systems: An architectural design decision framework. In: 2023 IEEE 20th International Conference on Software Architecture Companion (ICSA-C). IEEE (2023)

\bibitem{augur}
Lewis, G.A., Echeverría, S., Pons, L., Chrabaszcz, J.: Augur: A step towards realistic drift detection in production ml systems. In: 2022 IEEE/ACM 1st International Workshop on Software Engineering for Responsible Artificial Intelligence (SE4RAI) (2022)

\bibitem{industrial_automation}
Maschler, B., Weyrich, M.: Deep transfer learning for industrial automation: A review and discussion of new techniques for data-driven machine learning. IEEE Industrial Electronics Magazine  \textbf{15}(2) (2021)

\bibitem{healthcare}
Miotto, R., Wang, F., Wang, S., Jiang, X., Dudley, J.T.: Deep learning for healthcare: review, opportunities and challenges. Briefings in bioinformatics  \textbf{19}(6) (2018)

\bibitem{schwartz2020greenAI}
Schwartz, R., Dodge, J., Smith, N.A., Etzioni, O.: Green ai. Communications of the ACM  \textbf{63}(12) (2020)

\bibitem{shankar2020operationalizing}
Shankar, S., Garcia, R., Hellerstein, J.M., Parameswaran, A.G.: Operationalizing machine learning. University of California, Berkeley

\bibitem{maintainability2022slr}
Shivashankar, K., Martini, A.: Maintainability challenges in ml: A systematic literature review. In: 2022 48th Euromicro Conference on Software Engineering and Advanced Applications (SEAA) (2022)

\bibitem{energy_facebook}
Strubell, E., Ganesh, A., McCallum, A.: Energy and policy considerations for modern deep learning research. Proceedings of the AAAI Conference on Artificial Intelligence (09)

\bibitem{symeonidis2022mlops}
Symeonidis, G., Nerantzis, E., Kazakis, A., Papakostas, G.A.: Mlops - definitions, tools and challenges. In: 2022 IEEE 12th Annual Computing and Communication Workshop and Conference (CCWC) (2022)

\bibitem{tamburri2020sustainablemlops}
Tamburri, D.A.: Sustainable mlops: Trends and challenges. In: 2020 22nd International Symposium on Symbolic and Numeric Algorithms for Scientific Computing (SYNASC)

\bibitem{tedla2024ecomls}
Tedla, M., Kulkarni, S., Vaidhyanathan, K.: Ecomls: A self-adaptation approach for architecting green ml-enabled systems. arXiv preprint arXiv:2404.11411  (2024)

\bibitem{data_green_roberto}
Verdecchia, R., Cruz, L., Sallou, J., Lin, M., Wickenden, J., Hotellier, E.: Data-centric green ai an exploratory empirical study. In: 2022 International Conference on ICT for Sustainability (ICT4S) (2022)

\bibitem{wohlin2012experimentation}
Wohlin, C., Runeson, P., H{\"o}st, M., Ohlsson, M., Regnell, B., Wessl{\'e}n, A.: Experimentation in software engineering. Springer Science \& Business Media (2012)

\end{thebibliography}
